# Improved transport critical currents in Ag and Pb co-doped Ba$_x$K$_{1-x}$Fe$_2$As$_2$ superconducting tapes

Chao Yao[1], Chunlei Wang[1], Xianping Zhang[1], Lei Wang[1], Zhaoshun Gao[1], Dongliang Wang[1], Chengduo Wang[1], Yanpeng Qi[1], Yanwei Ma[1,*], Satoshi Awaji[2], and Kazuo Watanabe[2]

[1] Key Laboratory of Applied Superconductivity, Institute of Electrical Engineering, Chinese Academy of Sciences, PO Box 2703, Beijing 100190, China

[2] High Field Laboratory for Superconducting Materials, Institute for Materials Research, Tohoku University, Sendai 980-8577, Japan

*E-mail: ywma@mail.iee.ac.cn

**Abstract**

Fe-clad Ba$_x$K$_{1-x}$Fe$_2$As$_2$ superconducting tapes were fabricated by the *ex situ* powder-in-tube method combined with a short high-temperature annealing technique. The effect of annealing time and different dopants on the transport properties of the Ba$_x$K$_{1-x}$Fe$_2$As$_2$ tapes were systematically studied. By co-doping with Ag and Pb, the transport critical current density $J_c$ of Ba$_x$K$_{1-x}$Fe$_2$As$_2$ tapes was significantly improved in whole field region and the highest transport $J_c$ was up to $1.4 \times 10^4$ A/cm$^2$ ($I_c$=100 A) at 4.2K in self field. It is proposed that the superior $J_c$ in the co-doped samples are due to the combine effect of Pb doping at low fields and Ag doping at high fields.

---

*Author to whom any correspondence should be addressed.

## 1. Introduction

The discovery of superconductivity in LaFeAsO$_{1-x}$F$_x$ [1] in 2008 aroused enormous research on the iron-based superconductors. With high critical temperature [2-5] and upper critical field [6-8], the iron-based superconductors are very promising for high-field applications [9-10]. As an important part for practical applications, high-performance superconducting wires and tapes, which can be used in cables and magnets, need to be developed. Therefore, soon after the discovery of iron-based superconductors, (La, Sm) FeAsO$_{1-x}$F$_x$ (1111 type) [11-13], Sr$_x$K$_{1-x}$Fe$_2$As$_2$ (122 type) [14] and Fe(Se, Te) (11 type) [15] superconducting wires have been fabricated using the powder-in-tube (PIT) method. At the same time, some efforts such as sheath optimization and dopant addition have been made to attain and improve their transport current density $J_c$ [16]. However, in the polycrystalline iron-based superconductors, the misorientation of grains, inclusions at grain boundaries and other crystal defects strongly limit the transport current [17-19]. In order to improve the current carrying property, it is very important to obtain superconducting core with good grain connectivity, uniform composition, high density and low impurity content in iron-based superconducting wires and tapes.

At present, the optimal candidate for applications in iron-based superconductor family is the 122 type, which has relatively low anisotropy of 1-3, high pinning potential of $10^4$ K, moderate transition temperature, high critical field and critical current density [20]. In order to improve the transport $J_c$ of 122-type superconducting wires and tapes, Ag or Pb doped Sr$_x$K$_{1-x}$Fe$_2$As$_2$ wires and tapes were fabricated by our group [21, 22]. The results shows that the Pb doping can increase the transport $J_c$ in low field, but give no help in high field region. On the other hand, the Ag doping can effectively improve the transport J$_c$ in high field region. The $J_c$ enhancement by Ag doping was also proved by the recent report [23], which improved the $I_c$ of Ag doped Ba$_x$K$_{1-x}$Fe$_2$As$_2$ tape up to 60.7 A using a melting process at 1100 ℃ to prepare precursor.

Based on the previous work, this study aims to investigate the co-doping effects of Ag and Pb on the superconducting properties of Ba$_x$K$_{1-x}$Fe$_2$As$_2$ tapes. It is expected that by combining the advantages of the doping effect of Ag and Pb, the $J_c$ performance can be improved in the whole field region. For comparison, Ag and Pb doped Ba$_x$K$_{1-x}$Fe$_2$As$_2$ tapes are also fabricated following the same process. It is found that the transport $I_c$ of $Ba_xK_{1-x}Fe_2As_2$ tape at 4.2 K in self field is further increased to 100 A by co-doped with Ag and Pb, and its high-field performance is also improved compared to the case of single doping. The microstructure and superconducting properties of $Ba_xK_{1-x}Fe_2As_2$ tapes with single doping and co-doping were comparatively investigated.

**2. Experimental details**

The starting materials for preparing $Ba_{0.66}K_{0.48}Fe_2As_2$ precursor are small Ba pieces, K bulks and fine FeAs powder. They were mixed and then put in an $Al_2O_3$ crucible, which was immediately sealed in a stainless steel pipe by argon arc welding. Then the pipe was put into a tubular furnace keeping a temperature at 1100 ℃ for 10 minutes and cooled to room temperature by switching off the furnace. This heating temperature is higher than the melting point of FeAs compound, which can improve the mixing of constituent elements [23]. The precursor was ground into powder in an agate mortar and packed in an Nb tube, which was sealed in a stainless steel pipe and heat treated at 900 ℃ for 20 h. Then the sintered product was ground into fine powder and doped with Ag and Pb, following the stoichiometry $Ba_{0.66}K_{0.48}Fe_2As_2+Ag_{0.5}$, $Ba_{0.66}K_{0.48}Fe_2As_2+Pb_{0.2}$ and $Ba_{0.66}K_{0.48}Fe_2As_2+Ag_{0.5}+Pb_{0.2}$ respectively. Finally, the mixture was filled into Fe tubes with 8 mm outer and 5 mm inner diameter. These tubes were sequentially rotary swaged, drawn to wires, and rolled into tapes of 0.7 mm thick. In order to avoid the reaction between Fe sheath and superconducting core, a short high-temperature annealing technique [24] was used in the heat treatment of tapes. The tapes were cut into short samples of 6 cm long, which were sealed into Fe tubes and sintered at 1100 ℃ for 5, 10 and 15 minutes respectively. In the following parts of this paper, these samples are given their own code names for short, and the process parameters of them are summarized in table 1.

The phases of the superconducting core were studied by the x-ray diffraction (XRD) on a Rigaku D/MAX 2500 diffractometer. The microstructure of crushed superconducting core was analyzed with a Hitachi S4800 scanning electron microscope (SEM). The element distribution on the tape section was examined with energy dispersive x-ray spectroscopy (EDX) mapping and line scanning. The resistance measurements were carried out on a Quantum Design's physical property measurement system (PPMS) using four-probe method.

The transport critical current $I_c$ was measured at 4.2 K using short tape samples of 3 cm in length with standard four-probe method and evaluated by the criterion of 1 $\mu$V/cm, then the critical current was divided by the cross section area of the superconducting core to get the critical current density $J_c$. The applied fields in transport $I_c$ measurement are parallel to the tape surface.

## 3. Results and discussion

Figure 1(a) shows a typical SEM image of the cross section of $Ba_xK_{1-x}Fe_2As_2$ tapes prepared following the process mentioned above. We can see a clear boundary between the superconducting core and the Fe sheath, indicating that there is no obvious reaction at the inner surface of sheath during the heat treatment. This is essential for the transport $I_c$ measurement of the tapes [12, 25]. To further investigate the element distribution, EDX line-scan was performed on the cross section of undoped $Ba_xK_{1-x}Fe_2As_2$ tape. As shown in figure 1(b), most of the Ba, K and As elements are successfully restricted within the Fe sheath. For each element, the content has a dramatic change at the boundary of superconducting core. In figure 2, the EDX element mapping shows a similar result in Ag and Pb co-doped tape. Therefore, using *ex situ* PIT method combined with short high-temperature annealing technique, the reaction between superconducting core and Fe sheath can be avoided. In addition, the EDX mapping in figure 1(b) exhibits that the elements of parent compounds and dopants are homogeneously dispersed in superconducting core.

Compared to the previously reported bimetallic Ag/Fe sheath, using the Fe sheath can reduce the fabrication cost and simplify the structure of tapes, which is favorable for practical applications. More importantly, without the Ag inner sheath, the temperature of heat treatment can break the melting point of Ag (961.78 ℃), and the processing force during cold working can be applied directly on the superconducting core, which will promote its deformation and improve its density. As can be seen in figure 1, the superconducting core of tapes in this work looks denser than that in the Ag/Fe clad tape [21], which will is beneficial to the current carrying property of tapes.

The XRD patterns of several samples are given in figure 3. Except for some minor impurity peaks, the diffraction patterns show strong peaks of Ba-122 phase, indicating the main phase of all the samples is $Ba_{1-x}K_xFe_2As_2$ phase. The Ag or Pb peak can be respectively detected in the tapes of single doping, and both of them were simultaneously observed in the XRD pattern of co-doped sample. The Fe impurity detected in all the samples was introduced by the Fe sheath of tapes. On the other hand, it should be noted that the XRD patterns of BaKFeAs-AgPb-5 and BaKFeAs-AgPb-15 are almost the same, suggesting that the time of high-temperature annealing has no remarkable effect on the phases of samples.

Figure 4 gives the temperature dependence of the resistivity normalized by resistivity at 300 K for samples with various dopants and annealing time. The onset superconducting critical temperature $T_c$ of BaKFeAs-Ag-5, BaKFeAs-Pb-5 and BaKFeAs-AgPb-5 tape is 33.8 K, 31.4 K and 33.7 K, respectively. The $T_c$ of Ag doped samples is higher than the others, meaning that Ag has a positive effect to the $T_c$ of samples. On the other hand, we can see the superconducting transition of BaKFeAs-AgPb-10 and BaKFeAs-AgPb-15 is a little sharper than BaKFeAs-AgPb-5. This indicates that a longer annealing time can improve the crystallinity of the superconducting core, but its influence on $T_c$ is weaker than chemical addition.

In order to further understand the co-doping effect of Ag and Pb on the superconducting property, the temperature dependent resistivity of BaKFeAs-Pb-5 and BaKFeAs-AgPb-5 in various magnetic fields ($B$ = 0, 1, 3, 5, 7 and 9 T) was studied. With the increase of magnetic fields, the superconducting transition of these two samples shifts towards lower temperature, and become less steep. Their upper critical field $H_{c2}$ and irreversibility field $H_{irr}$, which are estimated with the criteria of 90% and 10% of resistivity at normal state respectively, are shown in figure 5. The upper critical field at zero-temperature $H_{c2}(0)$ was calculated using the Werthamer-Helfand-Hohenberg (WHH) formula: $H_{c2}(0) = -0.693T_c(dH_{c2}/dT)$. For BaKFeAs-Pb-5 and BaKFeAs-AgPb-5 samples, taking $T_c$ = 28.5 K and 31.5 K, the $H_{c2}(0)$ is 124 T and 129 T. In practical application, $H_{irr}$ is an important parameter which has a strong relationship with critical current density $J_c$. In figure 5, it can be seen that the $H_{irr}$ of BaKFeAs-AgPb-5 is much higher than BaKFeAs-Pb-5 at the same temperature, indicating that compared to the Pb doping, the Ag-Pb co-doping can further improve the superconducting property of $Ba_xK_{1-x}Fe_2As_2$ tapes when applying external magnetic field.

The transport critical $I_c$ of tapes with various dopants and annealing time in self field at 4.2 K is summarized in table 1. In general, the $I_c$ decreases with increasing annealing time, and the $I_c$ of co-doped tapes is about twice as much as the Pb doped samples with the same annealing time. Therefore, Ag-Pb co-doping can enhance the transport property in self field. Meanwhile, we can see that the $I_c$ of Ag doped tapes is much lower than the Pb doped and Ag-Pb co-doped tapes. This is different from the previous report, which can improve $J_c$ effectively by Ag doping only [21-23]. In their work, the tapes and wires were sintered at 850 - 900 ℃, whereas the tapes in this work were heated at 1100 ℃ after cold work. Hence the Ag dopant in $Ba_xK_{1-x}Fe_2As_2$ tapes may be sensitive to this change of the heat treatment condition.

The field dependent $J_c$ at 4.2 K of Pb doped and Ag-Pb co-doped tapes is given in figure 6. We can see that the $J_c$ of Pb doped tapes drop dramatically in increasing field. In contrast, the Ag-Pb co-doping can effectively enhance the $J_c$ performance in whole field region. In this work, the $J_c$ of the BaKFeAs-AgPb-5 tape achieves $1.4 \times 10^4$ A/cm$^2$ in self field. However, even in the co-doped sample, the $J_c$ still show a drop of about one order of magnitude in the field of 0.5 T. This is similar to the $J_c$ of cobalt-doped $BaFe_2As_2$ bicrystal films with large misorientation angles [19], indicating a characteristic of weak link in these tape samples. The inset of figure 6 shows the $J_c$ curves of BaKFeAs-AgPb-5 tape measured in increasing and decreasing fields. The hysteretic effect of $J_c$ is relevant to penetration of flux into strong pinning intragranular regions, and the presence of intragranular $I_c$ can enhance intergranular $I_c$ when the field is decreasing [26]. This also suggests that there are weak-linked current paths between grains in these tapes.

To investigate the effect of different dopants and annealing time on the tapes during the short high-temperature annealing of tapes, the SEM images of superconducting cores are shown in figure 7. In figure 7(a), there are some large Ag particles (marked by arrows) embedding in the matrix in BaKFeAs-Ag-5. These inhomogeneously dispersed Ag particles may be caused by the melting and segregation of Ag at the temperature as high as 1100 ℃. This is the reason why the Ag doping causes no remarkable improvement on the tapes under this short high-temperature annealing. In figure 7(b), we can see that the Pb additions (marked by arrows) bind the $Ba_xK_{1-x}FeAs$ matrix tightly in BaKFeAs-Pb-5, but their distribution in parent compound is also not very homogeneous. In the BaKFeAs-AgPb-5 sample of figure 7(c), we can see that the $Ba_xK_{1-x}FeAs$ particles and doping induced particles are finer and more uniform compared to the case of single doping, and it exhibits a better crystallinity than BaKFeAs-Pb-5 due to the Ag addition. Some very small voids (marked by circles) and a few impurity particles (marked by arrows) can be observed in this sample, but these defects are much smaller than those in BaKFeAs-Ag-5 and BaKFeAs-Pb-5. Accordingly, it seems that Pb doping can improve the particle connections, while Ag doping is helpful for grain formation. This is in accordance with the result in the tapes prepared with long time final annealing [27, 28]. When comparing figure 7(c) and (d), we can see that with the increase of annealing time, the egg-shape particles gradually melt into large irregular bulks in BaKFeAs-AgPb-15, which may lead to a relatively bad crystallinity.

A homogeneous microstructure with uniformly distributed dopants is very critical to achieve high superconducting properties in $Ba_xK_{1-x}Fe_2As_2$. Furthermore, the tapes with the high-temperature annealing less than 5 min will have a better $J_c$ performance, while the doping amount of the Ag-Pb co-doped tape still needs to be optimized. One the other hand, though the co-doping in this work can increase the $J_c$ in self field and the $J_c$ plateau in high field region, it do not eliminate the rapid drop of $J_c$ in low filed. Therefore, besides reducing defects such as pores, cracks, and inhomogeneous phases in $Ba_xK_{1-x}Fe_2As_2$ tapes, solve the weak link problem is another important issue to further improve the critical current in future.

## 4. Conclusions

In summary, we have fabricated Fe-clad $Ba_xK_{1-x}Fe_2As_2$ superconducting tapes by the *ex situ* powder-in-tube method combined with a short high-temperature annealing as the final heat treatment after the cold deformation. EDX mapping and line-scan shows that there is no reaction layer between the Fe sheath and the superconducting core. With Ag-Pb co-doping, the transport $J_c$ of tape was effectively enhanced in the whole field region. It seems that Pb doping can improve the particle connections, while Ag doping is beneficial to grain growth. Compared to Ag and Pb single addition, the Ag-Pb co-doping successfully combine their advantages, thus to achieve a further improvement of transport $J_c$.

**Acknowledgments**

This work is partially supported by the National Natural Science Foundation of China (grant Nos. 51025726 and 51104136), the National '973' Program (grant No. 2011CBA00105), and the Beijing Municipal Natural Science Foundation (grant no. 2122056).

Table 1. Summary of the information of $Ba_xK_{1-x}Fe_2As_2$ tape samples

| Stoichiometry | Sample Name | Annealing Time (min) | $T_{c,\ onset}$ (K) | Transport $I_c$ (A) (4.2 K 0 T) |
|---|---|---|---|---|
| $Ba_{0.66}K_{0.48}Fe_2As_2+Ag_{0.5}$ | Ag-5 | 5 | 33.8 | 1.5 |
| | Ag-10 | 10 | 33.8 | 1.6 |
| | Ag-15 | 15 | 33.9 | 1.9 |
| $Ba_{0.66}K_{0.48}Fe_2As_2+Pb_{0.2}$ | Pb-5 | 5 | 31.4 | 45 |
| | Pb-10 | 10 | 31.1 | 30 |
| | Pb-15 | 15 | 31.2 | 33 |
| $Ba_{0.66}K_{0.48}Fe_2As_2+Ag_{0.5}+Pb_{0.2}$ | AgPb-5 | 5 | 33.7 | 100 |
| | AgPb-10 | 10 | 34.0 | 83 |
| | AgPb-15 | 15 | 33.9 | 59 |

The heat treatment temperature of all the samples is 1100 $^o$C.

**Captions**

Figure 1 (a) SEM image of the cross section of $Ba_xK_{1-x}Fe_2As_2$ tape. The superconducting core and the Fe sheath are indicated by arrows. (b) EDX line-scan on the cross section of undoped tape. The scanning direction is indicated by the white solid line. From top to bottom, the color curves crossing the white dashed line represent intensities of Fe, Ba, K, O and As elements respectively.

Figure 2 EDX mapping images on the cross section of Ag and Pb co-doped tape.

Figure 3 XRD patterns of samples measured after grounding the superconducting core into powders. The peaks of impurity phases are marked with different symbols.

Figure 4 Temperature dependence of resistivity normalized by resistivity at 300 K for the tapes with various dopants and annealing time. The inset gives an expanded view at low temperature.

Figure 5 Temperature dependence of the upper critical field $H_{c2}$ and irreversibility field $H_{irr}$ for the BaKFeAs-Pb-5 and BaKFeAs-AgPb-5 tapes.

Figure 6 Transport critical current density $J_c$ of Pb doped and Ag-Pb co-doped tapes at 4.2 K in increasing fields. The inset shows the $J_c$ of BaKFeAs-AgPb-5 tape measured in increasing and decreasing fields successively.

Figure 7 SEM images of the superconducting cores of (a) BaKFeAs-Ag-5, (b) BaKFeAs-Pb-5, (c) BaKFeAs-AgPb-5 and (d) BaKFeAs-AgPb-15.

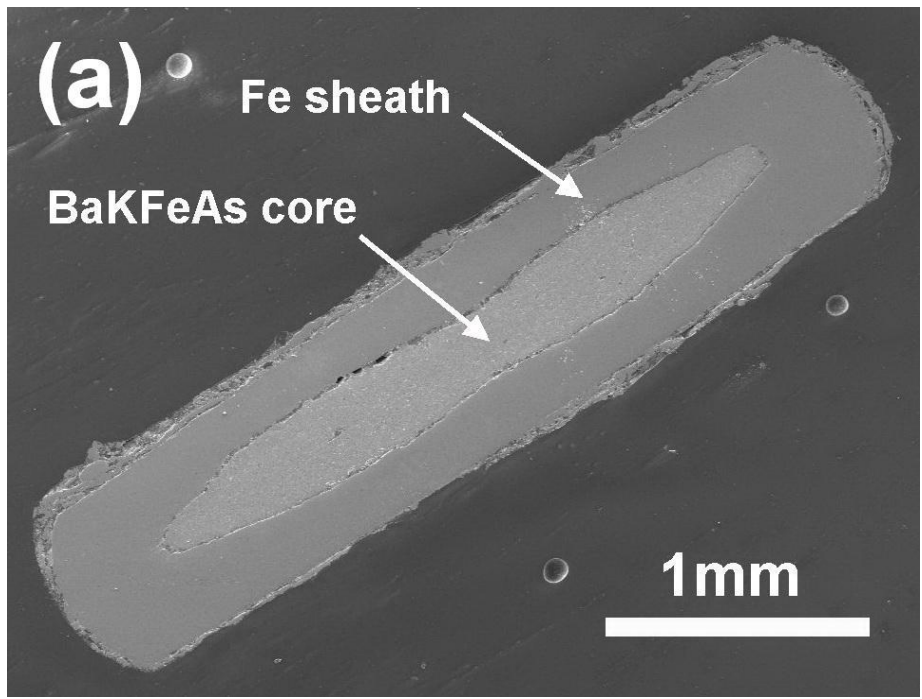

Figure 1(a) Yao et al.

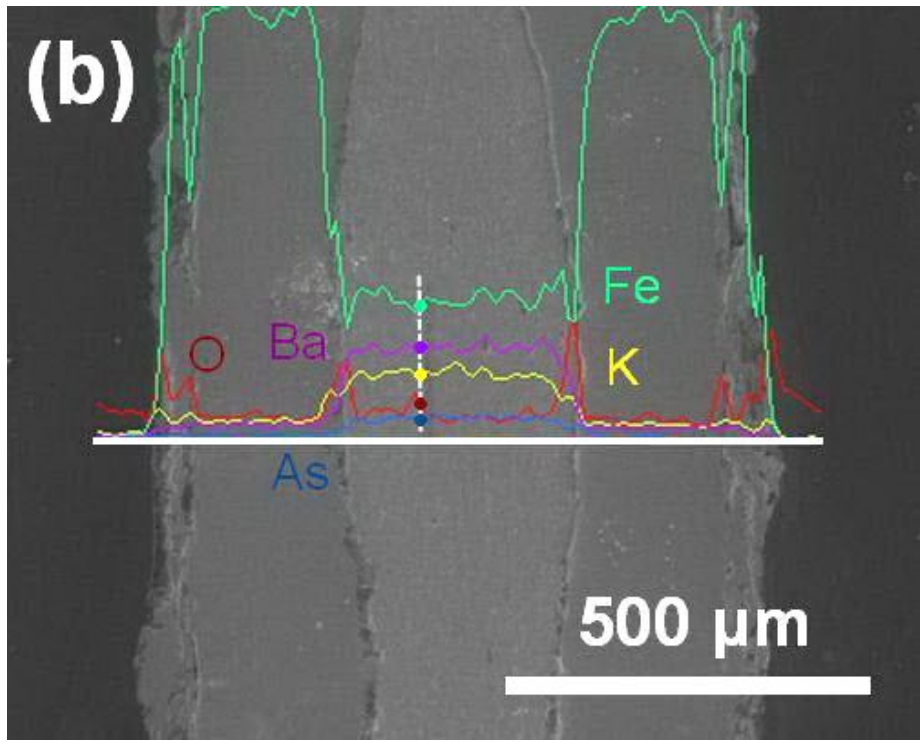

Figure 1(b) Yao et al.

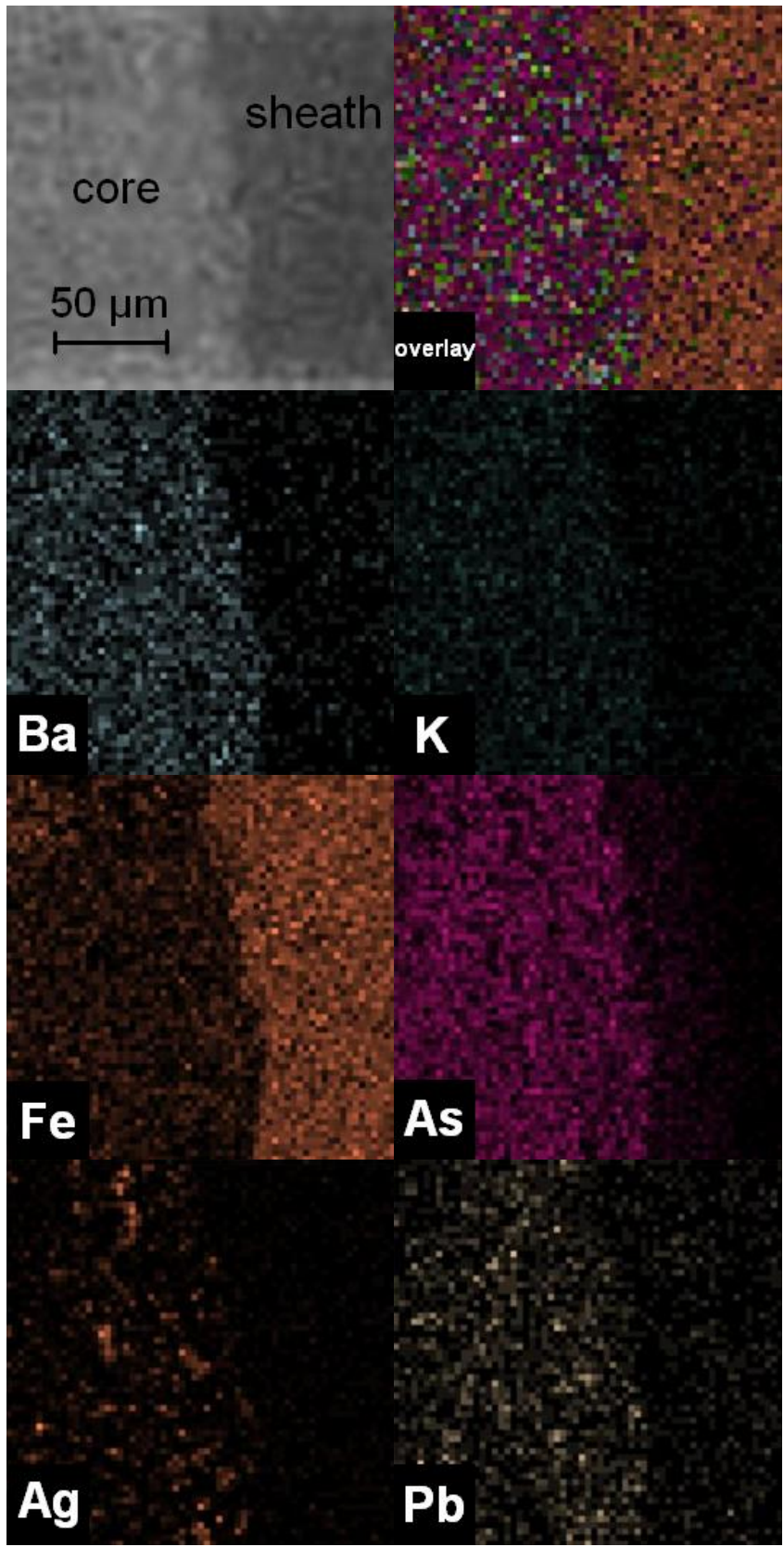

Figure 2 Yao et al.

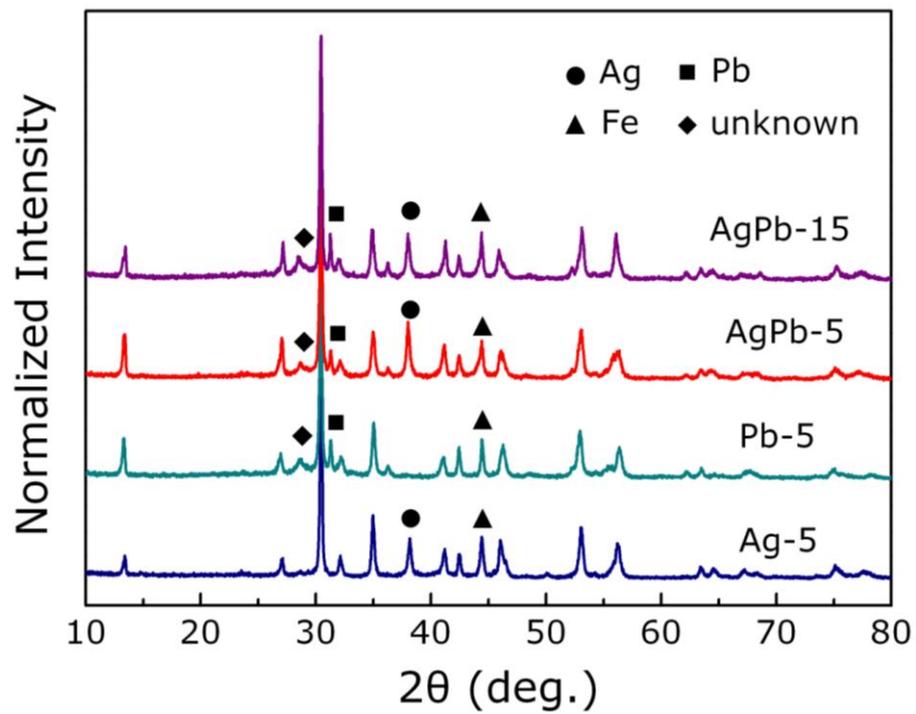

Figure 3 Yao et al.

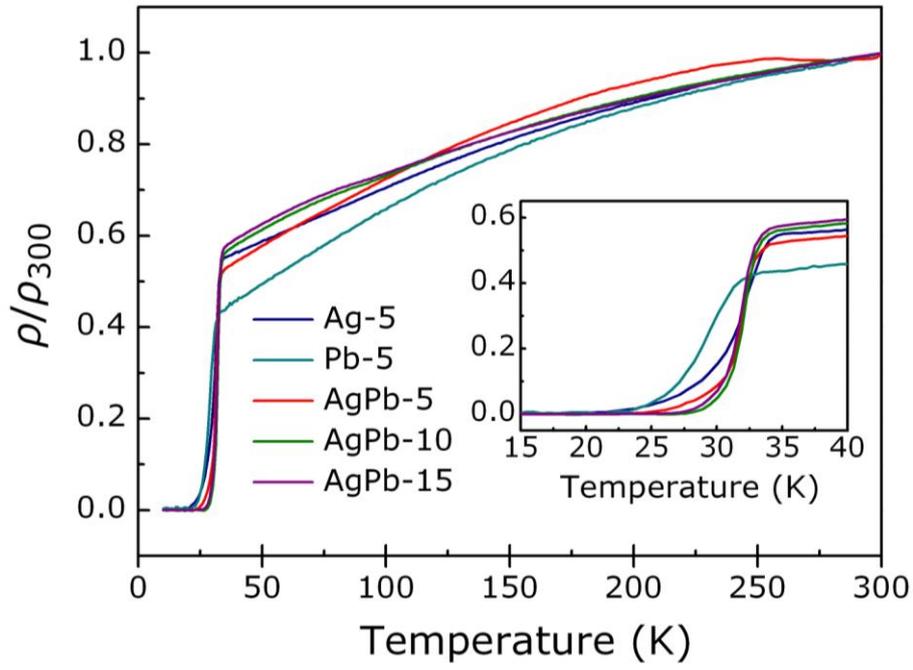

Figure 4 Yao et al.

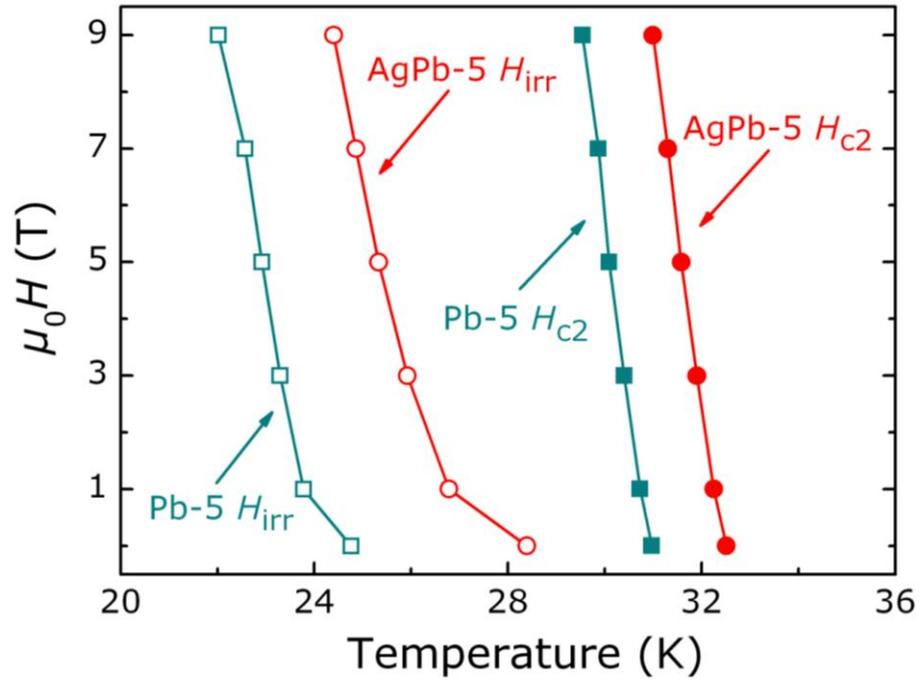

Figure 5 Yao et al.

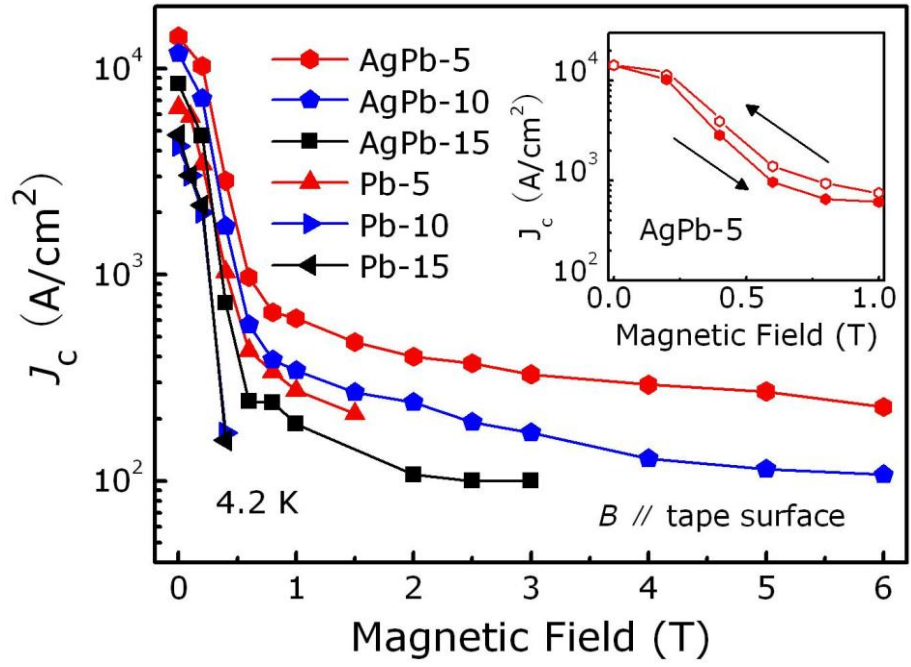

Figure 6 Yao et al.

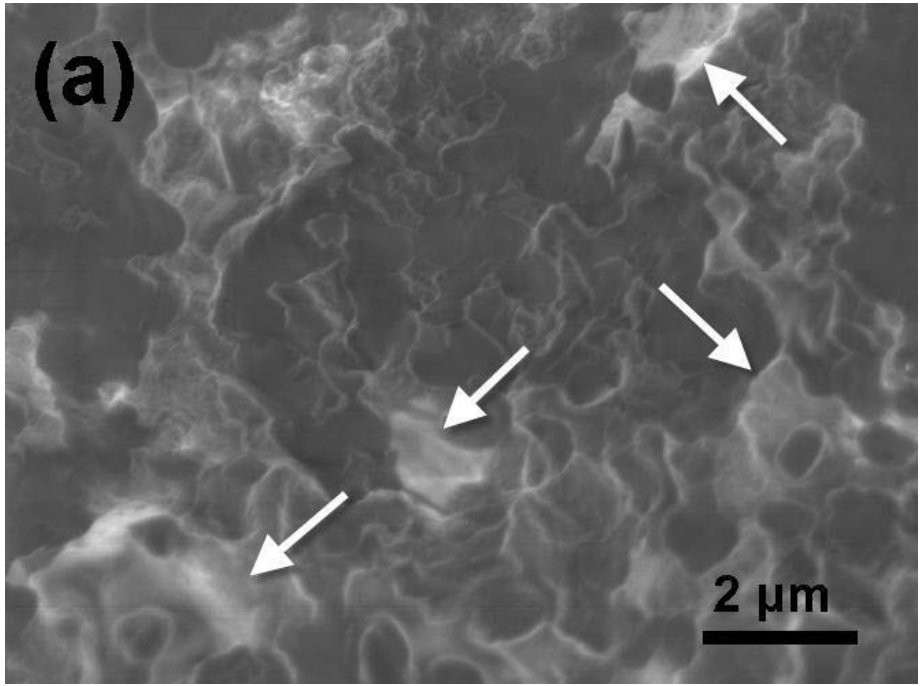

Figure 7(a) Yao et al.

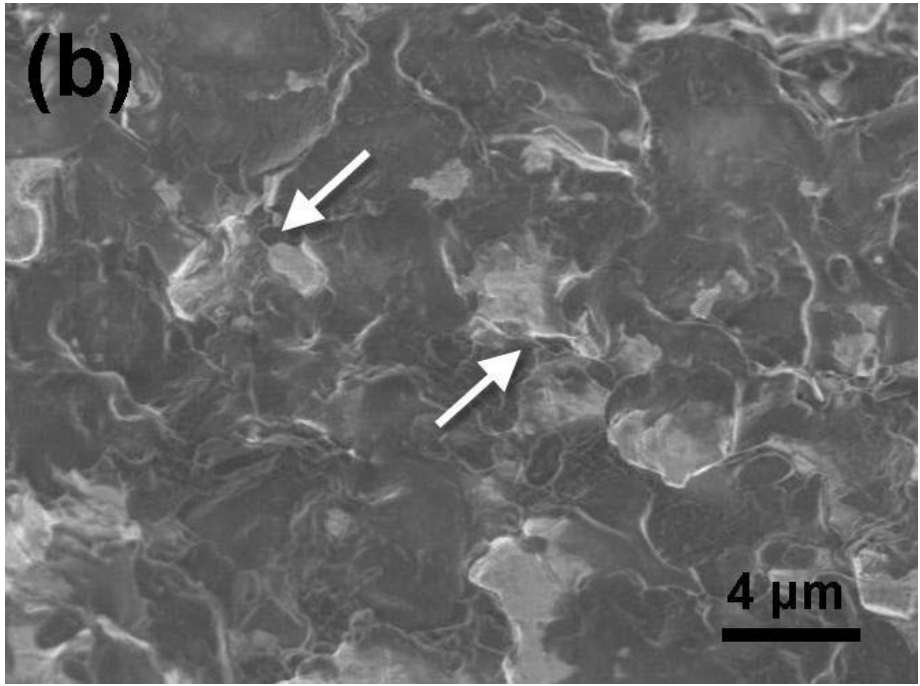

Figure 7(b) Yao et al.

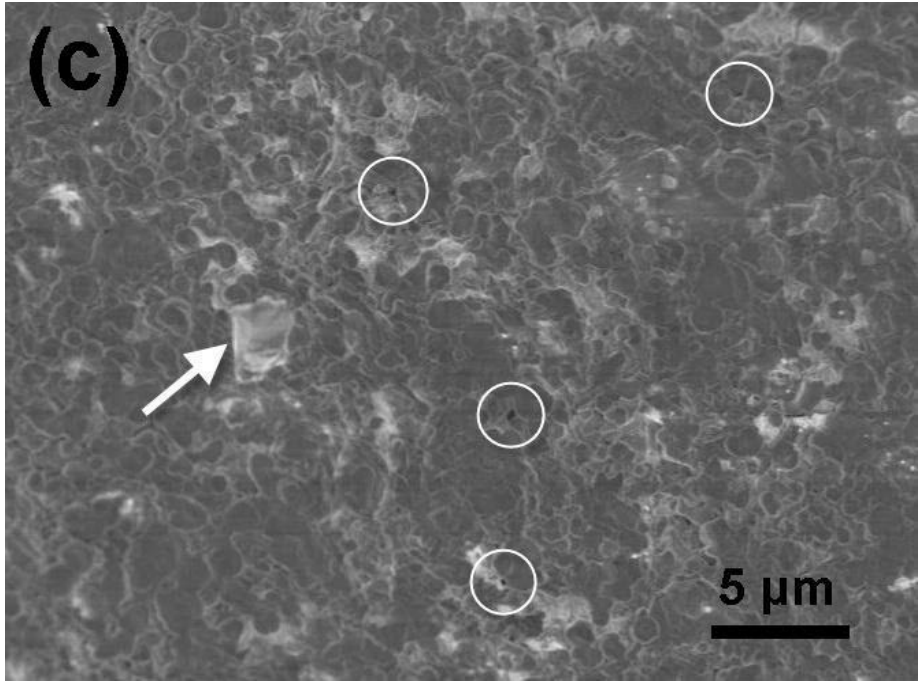

Figure 7(c) Yao et al.

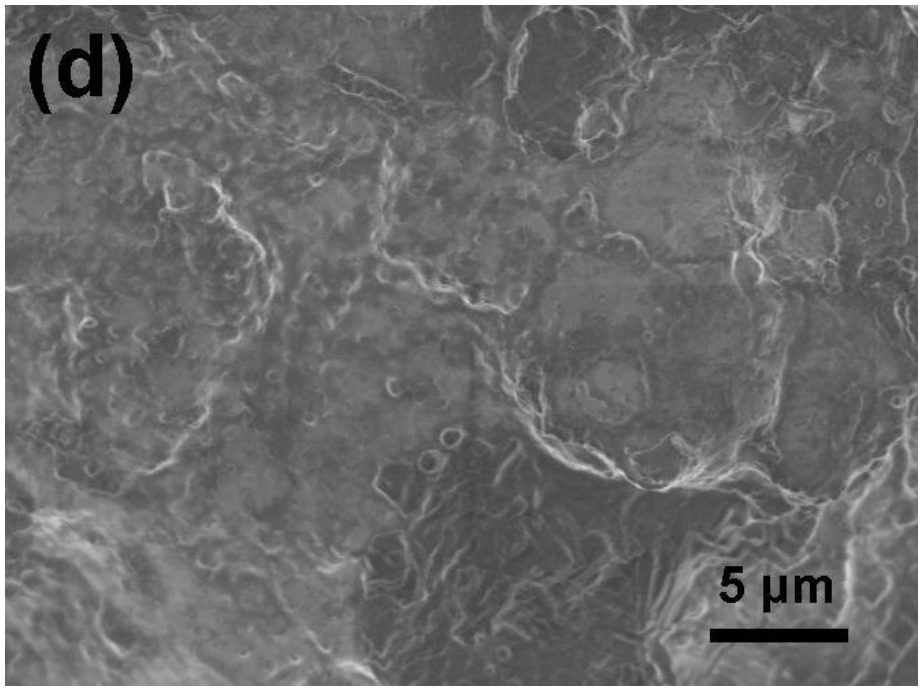

Figure 7(d) Yao et al.